
%
\input PHYZZX

%
\font\largemath = cmti10 scaled \magstep2

\textfont4=\largemath
\mathchardef\lS="0453
\def\pa{\partial}
\def\la{\langle}
\def\ra{\rangle_{\eta}}
\def\x{x^{\prime}}
\def\t{t^{\prime}}
\def\p{\phi_i^{\prime}}

\def\KO{[}
\def\KC{]}

\def\MCO{\Bigl\{}
\def\MCC{\Bigr\}}
\def\CO{\{}
\def\CC{\}}


\REF\pw{G.Parisi and Y.Wu, Sci.Sin. {\bf 24} (1981) 483}
\REF\dh{For a review, P.H.Damgaard and H.H\"uffel, Phys.Rep. {\bf 152} (1987)
227}
\REF\im{M. Ito and K. Morita, Prog.Theor.Phys. {\bf 87} (1992) 207 and
references there in}
\REF\njl{Y.Nambu and G.Jona-Lasinio, Phys. Rev. {\bf 122} (1961) 345}
\REF\ikm{K.Ikegami, T.Kimura and R.Mochizuki, CHIBA-EP-57 (1992)}
\REF\hig{K.Higashijima, Prog.Theor.Phys.Suppl. {\bf 104} (1991) 1}
\REF\imy{K.Ikegami, R.Mochizuki and K.Yoshida, CHIBA-EP-64 (1992)}
\REF\mochimochi{R.Mochizuki, CHIBA-EP-58 (1992)}
%
%
\unnumberedchapters
\date={}
%
\pubnum={CHIBA-EP-63-REV}
%
\date={September 1992}
\titlepage
\title{Large N Limit on  Langevin Equation: \break
Two-Dimensional Nonlinear Sigma Model}

\author{{\it Riuji} Mochizuki and {\it Kazuhiro} Yoshida }
\address{Department of Physics, Chiba University, \break
          1-33 Yayoi-cho, Chiba 260, Japan}

\baselineskip = 12pt

\baselineskip = 18pt
\section{Abstract}

We study the stochastic quantization of  two-dimensional nonlinear sigma
model in the large $N$ limit.  Our main tool is the {\it effective}
Langevin equation with which we investigate nonperturbative phenomena and
derive the results which are same as the path integral approach gives.
\vfill\eject

Recently, the stochastic quantization method\refmark\pw\refmark\dh has
been used to study spontaneous (dynamical) breakdown of
symmetry,\refmark\im for example,  Nambu$\cdot$Jona-Lasinio
model.\refmark\njl Their main tool for investigation is the stochastic
generating functional from which the effective potential is derived.  On
the contrary, the Langevin equation, which is very useful in perturbative
calculation, loses its position because the dynamical symmetry breaking
appears as a nonperturbative phenomenon.  Nevertheless if we can use it
even for nonperturbative calculation, we may expect that computation
will become easier and the relation between perturbative and
nonperturbative phenomena will become clearer.  In this paper we
introduce the effective Langevin equation for  two-dimensional
nonlinear sigma model in the large $N$ limit.  With the help of it we derive
some nonperturbative results which, of course, coincide with the results of the
path-integral quantization.

We consider the two-dimensional nonlinear sigma model
whose Lagrangian in Euclidean space-time is
$$
L={1\over 2}\pa_{\mu}\phi_i\pa_{\mu}\phi_i\eqno(1)
$$
$$
i=1,\cdots ,N\ \ ;\ \ \ \ \ \ \ \ \mu=1,2
$$
with the constraint
$$
\phi_i\phi_i={N\over g^2}.\eqno(2)
$$
Introducing a Lagrange multiplier field $\sigma$ and  taking account of the
constraint, we rewrite the Lagrangian as
$$
L_c={1\over 2}\pa_{\mu}\phi_i\pa_{\mu}\phi_i+{1\over
2}\sigma(\phi_i\phi_i-{N\over g^2}).\eqno(3)
$$

In the stochastic quantization scheme  fields depend on the
fictitious time $t$ and their development along the fictitious time
obeys the Langevin equation
$$
\dot \phi_i(x,t)=-{\pa L_c\over
\pa\phi_i}\mid_{\phi=\phi(x,t)}+\eta_i(x,t),\eqno(4)
$$
where a dot denotes  fictitious-time derivative and $\eta_i(x,t)$
is a noise field whose correlations are
$$
\la \eta_i(x,t)\ra=0,\eqno(5.a)
$$
$$
\la\eta_i(x,t)\eta_j(\x,\t)\ra=2\delta_{ij}\delta(x-\x)\delta(t-\t).
\eqno(5.b)
$$
Here the average manipulation $\la\cdots\ra$ is defined as
$$
\la\cdots\ra\equiv{1\over Z}\int D\eta(\cdots)\exp\MCO-{1\over 4}\int
d^2xdt\eta_i(x,t)\eta_i(x,t)\MCC,\eqno(6.a)
$$
$$
Z\equiv\int D\eta\exp\MCO-{1\over 4}\int
d^2xdt\eta_i(x,t)\eta_i(x,t)\MCC.\eqno(6.b)
$$

The Langevin equation for the field $\phi_i$ with the Lagrangian (3) becomes
$$
\dot\phi_i(x,t)=\pa^2\phi_i(x,t)-\sigma(x,t)\phi_i(x,t)+\eta_i(x,t).
\eqno(7)
$$
The equation (7) is studied perturbatively.\refmark\ikm In that case the
fields fluctuate around a point on the ($N-1$)-dimensional sphere
($\phi_i\phi_i={N\over
g^2}$), so that the vacuum loses the original O($N$) symmetry
and, as a result, ($N-1$) independent $\phi$'s become massless
Nambu-Goldstone bosons.  Nevertheless these particles suffer from
infrared divergence in two dimensional space time.  On the other hand it is
well-known that if we study the model nonperturbatively in the large $N$ limit
by ordinary quantization methods, the O($N$) symmetry recovers and the  field
gains a dynamical mass.\refmark\hig  In the following we introduce
the effective Langevin equation for the two-dimensional nonlinear
sigma model in the large $N$ limit and investigate nonperturbative
phenomena.  The method used here  is applicable to other models with
dynamical symmetry breaking.\refmark\imy

To evaluate $\sigma(x,t)$ in the Langevin equation (7) we regard
it as a functional of $\phi_i(x,t)$ and insert the unity
$$
1=\int
D\phi\mid J\mid\delta\CO\dot\phi_i(x,t)-\pa^2\phi_i(x,t)
+\sigma(x,t)\phi_i(x,t)-\eta_i(x,t)\CC
$$
into the generating functional (6.b) and integrate over the noise field
$\eta$.  Here $\mid J\mid$ denotes the Jacobian factor which is a divergent
constant\footnote{\dag}{We use Ito's calculation rule in this paper.}.  We
obtain
$$
Z=\int D\phi\mid J\mid\exp\MCO -{1\over 2}\int
d^2xdt {\bf L}\MCC,\eqno(8.a)
$$
where
$$
{\bf L}={1\over 2}(\dot\phi_i(x,t)-\pa^2\phi_i(x,t)
+\sigma(x,t)\phi_i(x,t))(\dot\phi_i(x,t)-\pa^2\phi_i(x,t)
+\sigma(x,t)\phi_i(x,t)).\eqno(8.b)
$$
We thus derive a field equation for $\sigma(x,t)$
$$
{\partial{\bf
L}\over\partial\sigma}=\phi_i(\dot\phi_i-\pa^2\phi_i+\sigma\phi_i)=0.\eqno(9)
$$
 Restoring the noise fields as the integration valiables of the
generating functional (8), we can obtain the Langevin equation (7)
again,\refmark\mochimochi in which $\sigma(x,t)$ is the solution of the
field equation (9)
$$
\sigma(x,t)=-{g^2\over
N}\phi_i(x,t)(\dot\phi_i(x,t)-\pa^2\phi_i(x,t)),\eqno(10)
$$
where the constraint (2) is taken into account.  In large $N$ expansion,
since $\delta_{ii}=N$, we may expand $\sigma(x,t)$  as
$$
\sigma(x,t)=\la\sigma\ra(t)+O(1/N)+\cdots.\eqno(11)
$$
To observe dependence of $\la\sigma\ra$ on the stochastic
time, we call the  Langevin equation for it;
$$
\eqalign{
\dot\sigma(x,t)=&-{\pa L_c\over\pa\sigma}\mid_{\phi=\phi(x,t)}+\xi\cr
=&-(\phi_i(x,t)\phi_i(x,t)-{N\over g^2})+\xi .\cr}
$$
Noting that the expectation value of one noise should vanish and taking
account of the constraint (2), the expectation value of the above
Langevin equation vanish
$$
\la\dot\sigma\rangle_{\eta,\xi}=0,
$$
which holds at any fictitious time.  Consequently $\la\sigma\ra$ does not
depend on the fictitious time and we will evaluate it at
$t\rightarrow\infty$ later.  In the following we study the Langevin
equation (7) exclusively in the large $N$ limit $N\rightarrow\infty$, which
enables us to treat it nonperturbatively since the Langevin equation
(7) no longer has any interaction terms.

To investigate the dynamical breaking of the $O(N)$ symmetry, we
decompose $\phi_i(x,t)$ into the expectation value
$\la\phi(x,t)\ra\equiv\Phi_i(t)$, which is the order parameter of the
$O(N)$ symmetry, and a fluctuating field $\p(x,t)$.  Then the Langevin
equation (7) becomes
$$
\dot\p(x,t)=(\pa^2-\la\sigma\ra)\p(x,t)-\CO\dot\Phi_i(t)+\la\sigma\ra\Phi
_i(t)\CC+\eta_i(x,t).\eqno(12)
$$
$\langle\sigma\rangle_{\eta}$ plays a role of square of dynamical mass in
the equation (12).  By definition of $\p(k=0,t)$, its expectation value must
vanish and consequently
   $$
\dot\Phi_i(t)=-\la\sigma\ra\Phi_i(t),\eqno(13)
 $$
which explicitly shows a relation between the order parameter $\Phi_i$ and
the square of the dynamical mass $\la\sigma\ra$.  If $\la\sigma\ra=0$, the
equation (13) imposes no restrictions on $\Phi_i$.  Nevertheless it is
easily known that the  massless bosons cause infrared divergence.  On the
other hand if $\la\sigma\ra>0$, $\Phi_i$ vanishes when
$t\rightarrow\infty$, namely, the $O(N)$ symmetry does not break down.

We easily obtain the solution of the equation (12);
$$
\p(k,t)=\int^t d\tau e^{
-(k^2+\la\sigma\ra)(t-\tau)}\MCO\eta_i(k,\tau)\MCC,\eqno(14)
$$
where
$$
\p(k,t)=\int{d^2x\over 2\pi}e^{ikx}\p(x,t),
$$
$$
\eta_i(k,t)=\int{d^2x\over 2\pi}e^{ikx}\eta_i(x,t).
$$
The correlations of $\eta_i(k,t)$'s are
$$
\la\eta_i(k,t)\ra=0,\eqno(15.a)
$$
$$
\la\eta_i(k,t)\eta_j(k^{\prime},\t)\ra=2\delta_{ij}
\delta(k+k^{\prime})\delta(t-\t).\eqno(15.b)
$$

Let us derive a gap equation and  compute
$\la\sigma\ra$.  Using the relations (10) and (13), $\la\sigma\ra$ is
written as
$$
\eqalign{
\la\sigma\ra=&{g^2\over
N}\lim_{t\rightarrow\infty}\la\phi_i(x,t)(-\dot\phi_i(x,t)+\pa^2\phi_i(x,t))\ra\cr
=&{g^2\over N}\lim_{t\rightarrow\infty}\la\Phi_i(t)\la\sigma\ra\Phi_i(t)+
\p(x,t)\pa^2\p(x,t)+N\delta^2(0)\ra.\cr}\eqno(16)
$$
In the second equality\footnote{\ddag}{The equation (16) can be directly
derived from the equation (7).} we have taken account of $\la\p(k,t)\ra=0$ and
Ito's calculation rule
$$
0=\lim_{t\rightarrow\infty}\la\dot F\KO\phi\KC\ra
=\lim_{t\rightarrow\infty}\MCO{\delta F\over\delta\phi_i}\dot\phi_i+{\pa^2
F\over\pa\phi-i\pa\phi_i}\delta(0)\MCC.
$$
We can easily
compute the last term with the equations (14) and (15)

$$
\eqalign{
{g^2\over N}\lim_{t\rightarrow\infty}\la\p(x,t)&\pa^2\p(x,t)\ra\cr =&{g^2\over
N}\int{d^2kd^2k^{\prime}\over(2\pi)^2}e^{-i(k+k^{\prime})x}
\lim_{t\rightarrow\infty}
\la-\p(k^{\prime},t)k^2\p(k,t)\ra\cr
=&-g^2\int{d^2k\over(2\pi)^2}{k^2\over k^2+\la\sigma\ra}\cr
=&{g^2\over 4\pi}\la\sigma\ra
ln({\Lambda^2\over\la\sigma\ra})-{g^2\Lambda^2\over 4\pi},\cr}
$$
where we have introduced a straight cutoff parameter $\Lambda$.  The two
divergent terms in the above equation and in the equation (16) cancel out.
Interested in non-zero solutions of $\la\sigma\ra$ we divide the both sides of
the equation (16) by it and  obtain the gap-equation
$$
1-{g^2\over N}\Phi_i(\infty)\Phi_i(\infty)={g^2\over
4\pi}ln({\Lambda^2\over\la\sigma\ra}).\eqno(17)
$$
Note that the above equation can be obtained by computing the expectation
value of $N/g^2=\la\phi_i(x,t)\phi_i(x,t)\ra$.

If we define the
renormalized coupling constant $g_r$ at a renormalization point $\mu$
as
$$
{1\over g_r^{\ 2}}\equiv{1\over g^2}-{1\over
4\pi}ln({\Lambda^2\over\mu^2}),
$$
the solution ot the gap-equation (17) is
$$
\la\sigma\ra=\mu^2\exp\CO-{4\pi\over g_r^{\
2}}+{4\pi\Phi_i(\infty)\Phi_i(\infty)\over N}\CC,\eqno(18)
$$
which is independent of the renormalization point $\mu$.  Remembering
the relation (13), we conclude that the $O(N)$ symmetry is not broken,
that is,
$$
\Phi_i(\infty)=0
$$
and the $n$ scalar particles gain a mass $m$
$$
m^2=\mu^2\exp\CO-{4\pi\over g_r^{\ 2}}\CC.
$$
These results are same as the path integral approach gives.

In this paper we have not discussed stability of the vacuum.
Nevertheless the methods developed here can be applied to, for example, the
Nambu$\cdot$Jona-Lasinio model\refmark\imy whose symmetry spontaneously
breaks down in the large N limit.  In that case the Langevin equations for
bound states are introduced and stability of the vacuum is considered
convergency of them.

\section{Acknowledgements}
We would like to thank Professor T. Kimura, Dr. A. Nakamura and  K.
Ikegami for useful discussions.   We also thank the referees for valuable
comments.

\vfill
\eject

\refout
\vfill
\eject


\end